\documentclass[runningheads]{llncs}

\usepackage{my_style}
\usepackage{joslyn}

\newcommand\blfootnote[1]{%
  \begingroup
  \renewcommand\thefootnote{}\footnote{#1}%
  \addtocounter{footnote}{-1}%
  \endgroup
}

\usepackage[T1]{fontenc}

\title{Topological Analysis of Temporal Hypergraphs}
\author{E
Audun Myers\inst{1} \and
Cliff Joslyn\inst{1} \and
Bill Kay\inst{2} \and
Emilie Purvine\inst{1} \and
Gregory Roek\inst{1} \and
Madelyn Shapiro\inst{1}
}
\authorrunning{Myers et al.}
\institute{Pacific Northwest National Laboratory, Mathematics of Data Science \and
Pacific Northwest National Laboratory, Computational Mathematics}
\date{January 2023}

\begin{document}

\maketitle

\begin{abstract}
    %!TEX root = ..\main.tex
%-------------------------------
%*******************************

\label{sec:abstract}

In this work we study the topological properties of temporal hypergraphs.
Hypergraphs provide a higher dimensional generalization of a graph that is capable of capturing multi-way connections. As such, they have become an integral part of network science. A common use of hypergraphs is to model events as hyperedges in which the event can involve many elements as nodes. This provides a more complete picture of the event, which is not limited by the standard dyadic connections of a graph. However, a common attribution to events is temporal information as an interval for when the event occurred. Consequently, a temporal hypergraph is born, which accurately captures both the temporal information of events and their multi-way connections.
Common tools for studying these temporal hypergraphs typically capture changes in the underlying dynamics with summary statistics of snapshots sampled in a sliding window procedure. However, these tools do not characterize the evolution of hypergraph structure over time, nor do they provide insight on persistent components which are influential to the underlying system. To alleviate this need, we leverage zigzag persistence from the field of Topological Data Analysis (TDA) to study the change in topological structure of time-evolving hypergraphs. We apply our pipeline to both a cyber security and social network dataset and show how the topological structure of their temporal hypergraphs change and can be used to understand the underlying dynamics.
\blfootnote{Information release number: PNNL-SA-181478}
\end{abstract}

%!TEX root = ..\main.tex
%-------------------------------
%*******************************

\section{Introduction} \label{sec:introduction}

Complex networks are a natural tool for studying dynamical systems where elements of the system are modeled in a dyadic way and evolve over time. There are many real-world examples, such as 
social networks~\cite{Skyrms2000}, 
disease spread dynamics~\cite{Husein2019}, 
manufacturer-supplier networks~\cite{Xu2019b}, 
power grid networks~\cite{Schaefer2018}, and
transportation networks~\cite{DavidBoyce2012}.
The underlying complex dynamical systems driving these networks cause temporal changes to their structure, with connections and elements added and removed as the dynamical system changes. We can summarize this category of complex network as dynamical networks~\cite{Harary1997} where the resulting graph is a temporal graph with temporal attributes associated to each connection and/or element of the complex network.

While temporal networks are useful in understanding systems with dyadic relations between elements, the complex network is not always satisfactory for modeling the relationship between multiple entities~\cite{Estrada2006}. For data with multi-way relations that cannot be described by dyadic connections, hypergraphs capture richer information about community structure. For example, in Section~\ref{subsec:SNA} we explore a hypergraph built from Reddit data (PAPERCRANE~\cite{Papercrane2022}) on threads about COVID-19. A dyadic model, where an edge links two users if and only if they posted in the same thread, loses all information about thread size. In contrast, a hypergraph, where each thread is an edge and a user is in a thread if and only if they posted in that thread, retains the total structure of the data. In this way, hypergraph analytics are a powerful tool when higher order structure is of interest.  Some instances where hypergraphs have been useful include human gene sets~\cite{Joslyn2021,Feng2021} where genes interact in complex combinations, cyber data~\cite{Joslyn2021} with the domain name systems mapping multiple domains and IPs, and social networks with interactions between large groups~\cite{Estrada2006}. 

In many use cases, individual snapshots of a complex system are less important than analysis of how the system {\em changes}. Often,  these networks are further improved by modeling them as Temporal HyperGraphs (THG) in the same way as temporal graphs, with temporal attributes (e.g., intervals or times) associated to the multi-way connections and elements. Examples can be found in many settings, such as anomaly detection in automotive data (CAN bus)~\cite{hanselmann2020canet} and cybersecurity using the Operationally Transparent Cybersecurity data we consider in Section~\ref{subsec:CDA}~\cite{golczynski2021end}.

Many common tools for studying the characteristics of THGs are based on summary statistics of the underlying hypergraph snapshots. These statistics provide insight to dynamic changes in the structure of the underlying hypergraph. For example, in \cite{Cencetti2021}, Cencetti \textit{et al.} studied temporal social networks as hypergraphs and were able to measure the burstiness of multi-way social interactions using a burstiness statistic. 
While statistics such as this can be informative for change detection and insights into the system dynamics, they are lacking in their ability to interpret changes in the structure of the temporal hypergraph. 
Another approach for studying temporal hypergraphs is through visual analytics. In~\cite{Fischer2021}, a matrix based visual analytics tool was designed for temporal hypergraph analysis which provides insights into the dynamic changes of the hypergraph. However, visualization tools are naturally limited in their ability to be automatically interpreted and often require expertise to properly understand. 

What distinguishes hypergraphs from graphs is that hyperedges come not only in arbitrary sizes, but also connected into arbitrarily complex patterns. As such, they can actually have a complex mathematical topology\footnote{Notice we use ``topology'' here in the formal sense, as distinct from how this is used informally in graph applications to refer to connectivity patterns in networks.} as complex ``gluings'' of multi-dimensional objects which can have a complex shape and structure. Studying the topology of hypergraphs is a becoming an increasingly large area, frequently exploiting their representation as Abstract Simplicial Complexes (ASCs).

%Due to these limitations, we aim to investigate an automatic shape summary of the time evolution of THGs. A natural solution to this is topological data analysis.

The field of Topological Data Analysis (TDA)~\cite{Edelsbrunner2002,Zomorodian2004} aims to measure the shape of data. Namely, data is represented as an ASC, whose homology is then measured to detect overall topological shape, including components, holes, voids, and higher order structures. However, this often requires the choice of parameters to generate the ASC from the data, which is typically nontrivial. Another, more automatic, approach for measuring the shape of data is to use persistent homology from TDA. This method for studying the shape of data extracts a sequence of ASCs from the data, which is known as a filtration. Persistent homology has been successfully applied to a wide application domains, including manufacturing~\cite{Khasawneh2016,Yesilli2022}, 
biology~\cite{Amzquita2020}, 
dynamical systems~\cite{Myers2019,Tempelman2020}, 
and medicine~\cite{Skaf2022}. Many of the applications either represent the data as point clouds or graphs. For point cloud data, filtrations are commonly generated as a collection of  Vietoris-Rips complexes~\cite{Edelsbrunner2002} determined by identifying points within a Euclidean  distances of an increasing radius parameter. For graph data a similar process would be to use a distance filtration with the shortest path distance~\cite{Myers2019,Aktas2019}. 

Hypergraphs have also been studied using tools from TDA. Namely, the work in~\cite{Gasparovic2021} shows how the homology of hypergraphs can be studied using various ASC representations such as the associated ASC~\cite{Ren2020} or the relative/restricted barycentric subdivision.

However, a requirement for applying persistent homology is that there is a monotonic inclusion mapping between subsequent members of a sequence of ASCs (i.e., each subsequent ASC in the sequence has its previous as a subset). Many sequences of ASCs associated with data sets are not monotonic, however, we still want to track their changing structure. This is commonly true for temporal data, where, for example, hypergraph edges can appear and then disappear over time, which would break the monotonicity  requirement for persistent homology.

To solve this problem, zigzag persistence~\cite{Carlsson2010} can be applied. 
Instead of measuring the shape of static point cloud data through a distance filtration (e.g., a Vietoris-Rips filtration), zigzag persistence measures how long a topology generator persists in a sequence of ASCs by using a an alternating sequence of ASCs, called a ``zigzag filtration''. 

Both PH and zigzag persistence track the formation and disappearance of the homology through a persistence diagram or barcode as a two-dimensional summary consisting of persistence pairs $(b, d)$, where $b$ is the birth or formation time of a generator of a ``hole'' of a certain dimension, and $d$ is its death or disappearance time. For example,
in~\cite{Tymochko2020} the Hopf bifurcation is detected through zigzag persistence of Vietoris-Rips complexes over sliding windows using the one-dimensional homology. Another recent application~\cite{Myers2022a} studies temporal networks, where graph snapshots were represented as ASCs using the Vietoris-Rips complex with the shortest path distance.
However, both of these methods require a distance parameter to be selected to form the ASC at each step, which is typically not a trivial choice.

The resulting persistence barcodes from zigzag persistence can also be vectorized using methods such as persistence images~\cite{Adams2017} or persistence landscapes~\cite{Bubenik2015}. This allows for the resulting persistence diagrams to be analyzed in automatic methods using machine learning for classification or regression.

In this work we leverage zigzag persistence to study THGs. By measuring the changing structure of the temporal hypergraph through an ASC representation of the hypergraph, we are able to detect the formation, combination, and separation of components in the hypergraph as well as higher dimensional features such as loops or holes in the hypergraph and voids. The detection of these higher dimensional features is critical for temporal hypergraph analysis as they may be of consequence depending on the application domain. Additionally, in comparison to creating an abstract ASC from point cloud or graph data, no distance parameter needs to be chosen as there are natural representations of a hypergraph as an ASC~\cite{Gasparovic2021}.

In Section~\ref{sec:background} of this paper we  introduce THGs and an ASC representation of hypergraphs as well as an overview of zigzag persistence and how these are incorporated into our method of applying zigzag persistence for studying THGs. This section also includes a toy example demonstrating each step in the pipeline. In Section~\ref{sec:results} we demonstrate how our method can be applied to two data sets drawn from social networks and cyber data. Lastly, in Section~\ref{sec:conclusion} we provide conclusions and future work.

\section{Method and Background} \label{sec:background}
\begin{figure}[h!] 
    \centering
    \includegraphics[width=1\textwidth]{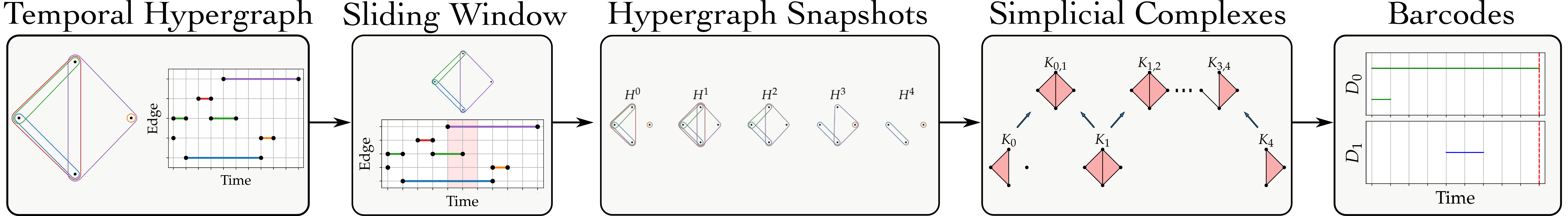}
    \caption{Pipeline for applying zigzag persistence to temporal hypergraphs.}
    \label{fig:pipeline_ZZ_THG}
\end{figure}

In this section the method for studying temporal hypergraphs using zigzag persistence is developed alongside the necessary background material.
Our method is a confluence of zigzag persistence and the ASC representation of hypergraphs for the topological analysis of THGs. Namely, we develop a pipeline for applying zigzag persistence to study changes in the shape of a temporal hypergraph using a sliding window procedure. This pipeline is outlined in Fig.~\ref{fig:pipeline_ZZ_THG}. 

We begin with a temporally edge-attributed hypergraph in Fig.~\ref{fig:pipeline_ZZ_THG}--\textit{Temporal Hypergraph}, where each edge has active intervals associated to it as described in Section~\ref{ssec:THG}. 
Next, we use a Fig.~\ref{fig:pipeline_ZZ_THG}--\textit{Sliding Window} procedure, where we choose a window size $w$ and shift $s$ that is slid along the time domain of the set of intervals in discrete steps. Using each sliding window, we generate Fig.~\ref{fig:pipeline_ZZ_THG}--\textit{Hypergraph Snapshots} at each window, which is described in Section~\ref{ssec:sliding_windows_for_HGs}. 
We then represent each snapshot as a Fig.~\ref{fig:pipeline_ZZ_THG}--\textit{ASC} using the associated ASC in Section~\ref{ssec:associated_SC}. 
Next, we introduce simplicial homology for studying the shape of an ASC in Section~\ref{ssec:simplicial_homology}.
This leads to the method for relating the homology within a sequence of ASCs known as zigzag persistent homology in Section~\ref{ssec:ZZ_THG}, which is used for calculating the persistent homology of the temporal hypergraph represented as a barcode of persistent diagram (Fig.~\ref{fig:pipeline_ZZ_THG}--\textit{Barcodes}).

To illustrate our procedure we provide a simple example throughout each step in the pipeline.
For the example and the remaining results we use the Python packages \texttt{HyperNetX}\footnote{HyperNetX: \text{https://pnnl.github.io/HyperNetX}} to generate the hypergraphs and \texttt{Dionysus2}\footnote{Dionysus2: \text{https://mrzv.org/software/dionysus2/}} to calculate the zigzag persistence. 

\subsection{Temporal Hypergraphs} \label{ssec:THG}
\begin{figure}[h!] 
    \centering
    \begin{subfigure}[t]{.23\textwidth}
        \centering
        \includegraphics[width=.8\textwidth]{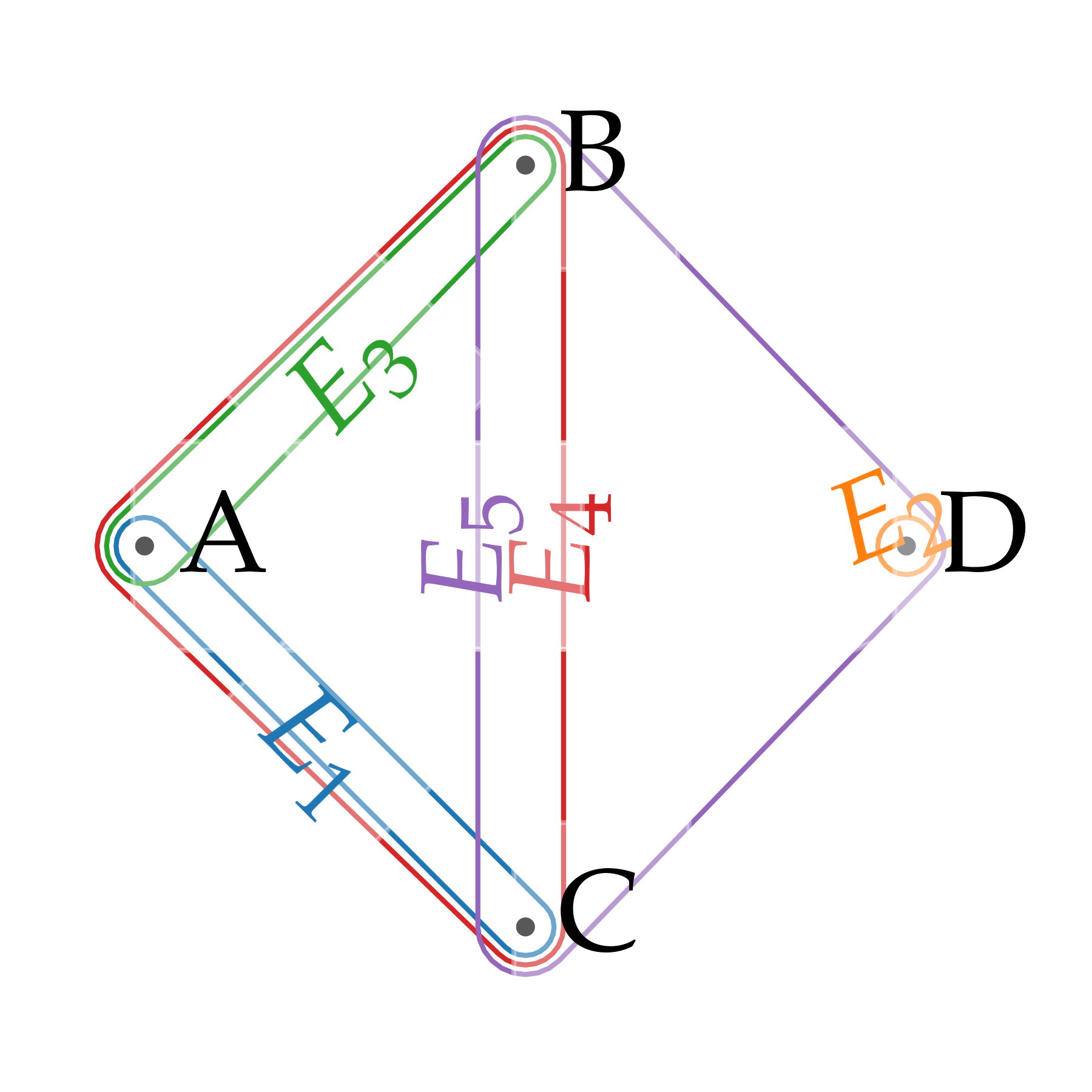}
        \caption{Static hypergraph}
        \label{fig:toy_example_THG_HG}
    \end{subfigure}
    \hspace{15pt}
    \begin{subfigure}[t]{.34\textwidth}
        \centering
        \includegraphics[width=.8\textwidth]{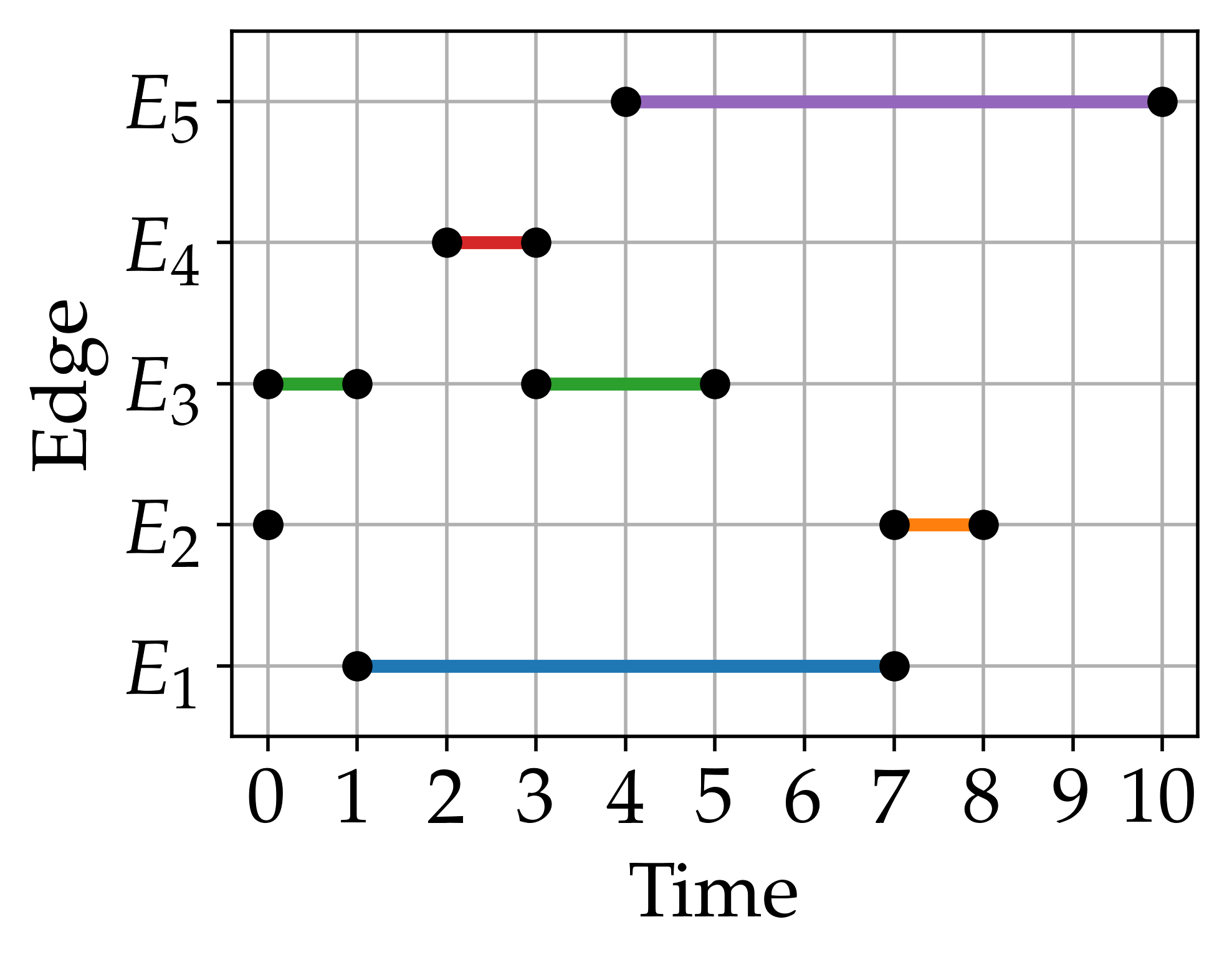}%
        \caption{Temporal information stored as points and intervals for each edge.}
        \label{fig:toy_example_THG_intervals}
    \end{subfigure}
    \caption{Toy example temporal hypergraph.}
    \label{fig:toy_example_THG}
\end{figure}

A graph $G ( V, E )$ is composed of a set of vertices connected using a set of edges with $ E \subseteq { V \choose 2}$. 
A hypergraph $H(V,E)$ is composed of a set of vertices $V$ and a family of edges $E$, where for each $E_i \in E, E_i \subseteq V$. In this way a hypergraph can capture a connection between $k$ vertices as a $k$-edge. For example, consider the toy hypergraph in Fig.~\ref{fig:toy_example_THG_HG} with four nodes $V = \{ A, B, C, D \}$ and five hyperedges $E = \{E_1, E_2, E_3, E_4, E_5\}$. These hyperedges in the example range in size from edge $E_2 = (D)$ as a 1-edge to edge $E_4 = (A, B, C)$ as a 3-edge.

A temporal hypergraph $H(V, E, T)$ is a replica of its underlying static hypergraph with the addition of temporal attributes $T$ associated to either the vertices, edges, or incidences. An attribute to an incidence occurs when the temporal information associated to a node is relative to the hyperedge. In this work we only use temporal information attributed to the edges. However, our pipeline could be adapted to any or all of the three temporal attribution types.
Returning to our toy example hypergraph $H$ in Fig.~\ref{fig:toy_example_THG_HG}, we include temporal information as a set of intervals associated to the time when each edge is active (e.g., $E_2$ is active for the point interval $[0,0]$ and interval $[7,8]$). 

\subsection{Sliding Windows for Hypergraph Snapshots} \label{ssec:sliding_windows_for_HGs}
The sliding window procedure is a ubiquitous part of signal processing, in which a time series or signal is segmented into discrete windows that slide along its time domain. Specifically, Given a time domain $[t_0, t_f]$, window size $w$, and shift $s$, we create a set of windows that cover the time domain interval as 
\begin{equation}
    \mathcal{W} = \{[t_0, t_0 + w], [t_0 + s, t_0 + s + w], [t_0 + 2s, t_0 + 2s + w], \ldots, [t_0 + \ell s, t_0 + \ell s + w]\},
    \label{eq:windows}
\end{equation}
The window size and shift should be such that $s \leq w$. In this way the union of all windows covers the entire domain and adjacent windows do not have a null intersection.

For each sliding window $W_i \in \mathcal{W}$ we create a sub-hypergraph snapshot using an intersection condition between the sliding window interval $W_i$ and the collection of intervals associated to each edge in the temporal hypergraph. The intervals are considered closed intervals in this work. This procedure is done by including an edge if there is a nonempty intersection between the edge's interval set and the sliding window interval $W_i$. We formalize this as
\begin{equation}
    H_i = \{E_j \in E \; | \; I(E_j) \cap W_i \neq \emptyset \},
\end{equation}
where $E_j \in E$ is an edge in the set of edges of the static hypergraph and $I(E_j)$ is the interval collection for edge $E_j$.
The resulting sub-hypergraph snapshot collection of $\mathcal{W}$ is $${\cal H} = \{ H_0, H_1, \ldots, H_t, \ldots, H_\ell \}.$$ 
We can cast this collection as a discrete dynamical process $H_t \mapsto H_{t+1}$ to gain understanding of the underlying system's dynamics. 

\begin{figure}[h!] 
    \centering
    \includegraphics[width=1\textwidth]{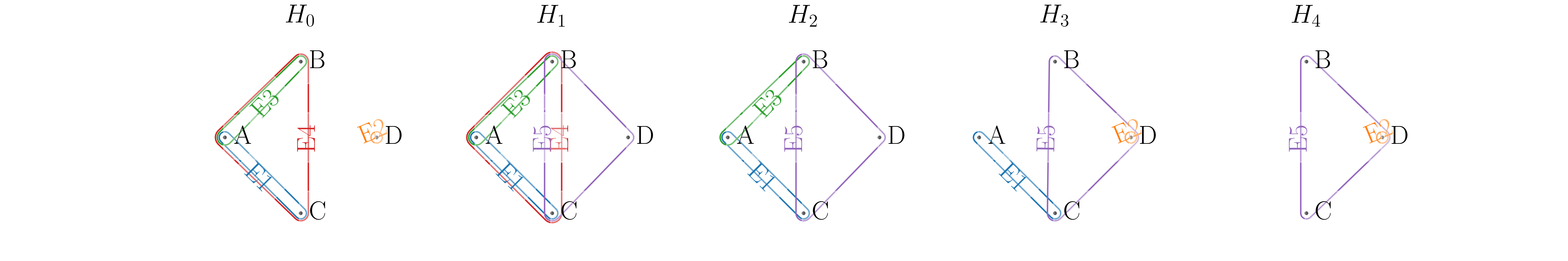}
    \caption{Sequence of sub-hypergraphs $\mathcal{H}$ from the sliding window procedure with corresponding ASCs.}
    \label{fig:H_THG_snapshots}
\end{figure}

To demonstrate the sliding window procedure for getting hypergraph snapshots we use the toy example temporal hypergraph from Fig.~\ref{fig:toy_example_THG} and window parameters $w = 2$ and $s=2$.
Using these parameters we get the sliding windows as $$\mathcal{W} = \{ [0,2], [2,4], [4,6], [6,8], [8,10] \}.$$ Hypergraphs from each window are generated as subsets of the static $H$ depending on the overlap of the window and the activity intervals associated to each edge. 
For example, window $W_2 = [4, 6]$ has the hypergraph $H_2$ with edges $\{ E_1, E_3, E_5 \}$ based on the overlap between $W_2$ and the collection of intervals of each edge shown in Fig.~\ref{fig:toy_example_THG_intervals}. 
Additionally, each hypergraph has now both an index and time associated to it. The index is as was previously stated (e.g., $H_2$ has index $2$) and the time is the average time of the corresponding window (e.g., $W_2$ has an average time of $(4+6)/2 = 5$). 
Applying this hypergraph snapshot procedure using the sliding windows we get the five hypergraphs shown in Fig.~\ref{fig:H_THG_snapshots}.

\subsection{Associated ASC of a Hypergraph} \label{ssec:associated_SC}

An ASC $K$ is a collection of simplices, with a simplex $\sigma \subseteq P$ as a subset of $n$ points from a set of points $P$ and simplex dimension $n-1$. This results in points (1-edge) as 0-simplices, lines (2-edge) as 1-simplices, triangles (3-edge) as 2-simplices, etc. We denote the simplex $\sigma$ as a face if $\sigma \subseteq \tau$ with $\tau$ as another simplex.
Additionally, a simplex $\sigma$ of dimension $n-1$ is required to be closed under face relation, which is all of its subsimplices (faces) as the power set of the simplex.
The dimension of an ASC is the dimension of the largest simplex.
ASCs are often used to represent geometric structures and as such are referred to as geometric simplicial complexes. However, we can also refer to them as abstract simplicial complexes for purely combinatorially purposes.

We can generate the associated ASC of a hypergraph~\cite{Ren2020} using the simplices associated to each hyperedge and building the closure under face relations, which is the power set of each hyperedge. 

To apply zigzag persistence to study the changing topology of our hypergraph snapshots, we need to first represent our collection of hypergraph snapshots $\mathcal{H}$ as a sequence of ASCs $\mathcal{K}$ which will later be used to create the zigzag persistence module. 
While there are several methods for representing a hypergraph as an ASC~\cite{Gasparovic2021}, we leverage an adaptation of the associated ASC method from~\cite{Ren2020}. 
The associated ASC of a hypergraph $H$ is defined as
\begin{equation}
    K(H) = \{ \sigma \in \mathcal{P}(E_i) \setminus \emptyset \; | \; E_i \in E\},
    \label{eq:original_associated_simplicial_complex}
\end{equation}
where $E$ is the edge set of the hypergraph $H$, $E_i \in E$, and $\mathcal{P}(E_i)$ is the power set of $E_i$. Equation~\ref{eq:original_associated_simplicial_complex} provides a first starting point for calculating the zigzag persistence, however, it is computationally cumbersome. Specifically, for a large $k$-edge the computational requires 
$$\sum_{j=0}^k {{k+1} \choose {j+1}} = 2^{k+1} - 1$$ 
subsimplices. However, the computation of homology of dimension $p$ only requires simplices of size $p+1$ to be included in the ASC. As such, we define the \textit{modified associated ASC} as
\begin{equation}
    K(H, p) = \{ \sigma \in \mathcal{P}_{p+1}(E_i) \setminus \emptyset  \; | \; E_i \in E\},
    \label{eq:modified_associated_simplicial_complex}
\end{equation}
where $\mathcal{P}_{p+1}$ is the modified power set to only include elements of the set up to size $p+1$ or ${E_i \choose {p+1}}$. The modified associated ASC reduces the computational demand by only requiring $$\sum_{j=0}^{p+1} {{k+1} \choose {j+1}}$$ subsimplices for a $k$-edge.

\begin{figure}[h!] 
    \centering
    \includegraphics[width=0.8\textwidth]{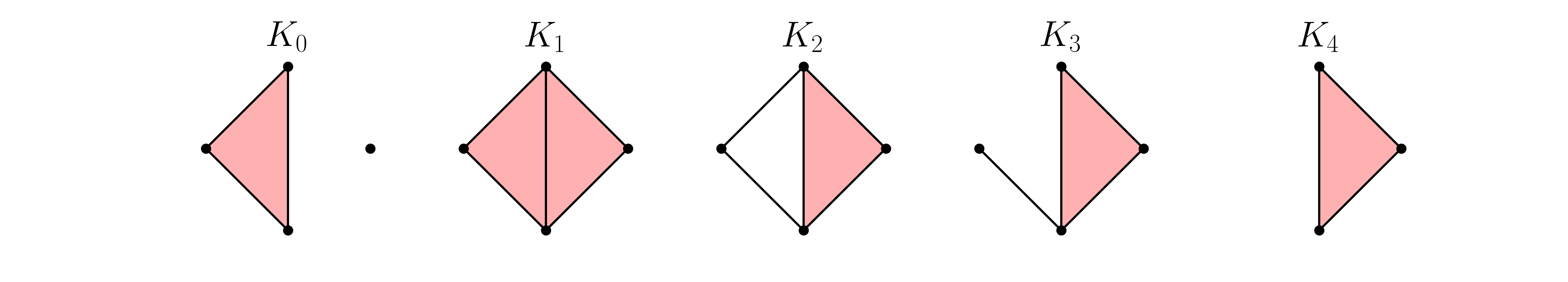}
    \caption{Sequence of associated ASCs from hypergraph snapshots in  Fig.~\ref{fig:H_THG_snapshots}.}
    \label{fig:K_THG__no_inter_snapshots}
\end{figure}

Applying Eq.~\eqref{eq:modified_associated_simplicial_complex} to each hypergraph in $\mathcal{H}$ allows us to get a corresponding sequence of ASCs as $\mathcal{K}$. For the hypergraph snapshots $\mathcal{H}$ shown in Fig.~\ref{fig:H_THG_snapshots} the modified associated ASCs $\mathcal{K}$ are shown in Fig.~\ref{fig:K_THG__no_inter_snapshots}.

\subsection{Simplicial Homology} \label{ssec:simplicial_homology}

Simplicial homology is an algebraic approach for studying the shape of an ASC by counting the number of $p$-dimensional holes, where $p=0$ are connected components, $p=1$ are graph triangles, $p=2$ are three-dimensional hollow tetrahedrons, and so on.
We can represent the collection of
$p$-dimensional holes of an ASC $K$ as the Betti vector $\beta(K) = [b_0, b_1, b_2, \ldots]$, where $b_p$ is the number of $p$-dimensional holes known as a Betti number. In this work we do not overview the details on how the Betti numbers are calculated, but we direct the reader to \cite{Otter2017,Munch2017} for a formal introduction.

\begin{figure}[h!] 
    \centering
    \includegraphics[width=0.7\textwidth]{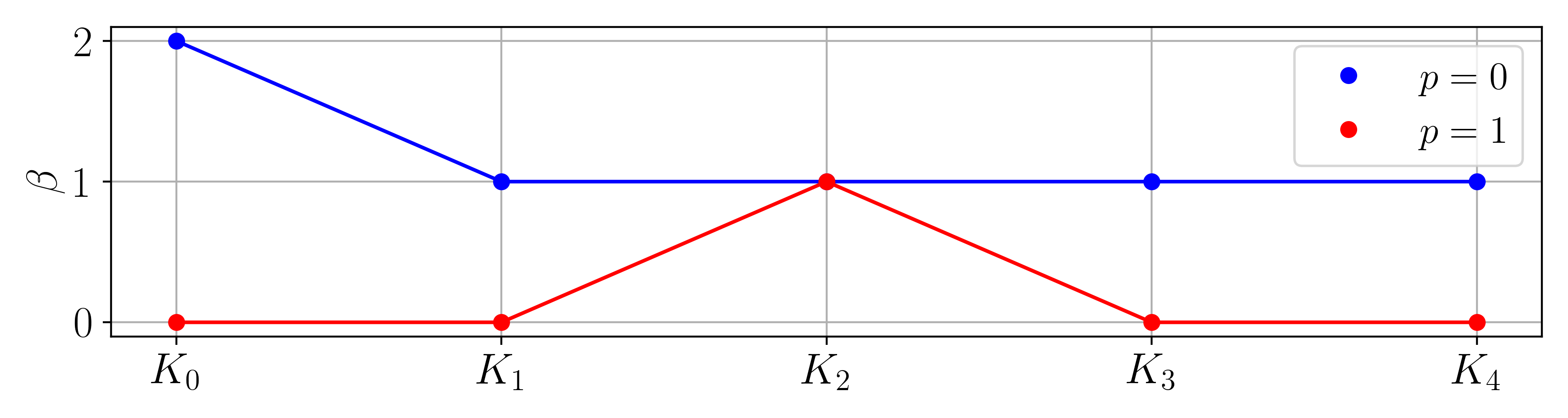}
    \caption{Betti numbers for ASCs in  Fig.~\ref{fig:K_THG__no_inter_snapshots}.}
    \label{fig:betti_numbers_toy_example}
\end{figure}

By calculating the Betti numbers for our sequence of ASCs in Fig.~\ref{fig:K_THG__no_inter_snapshots}, we get the Betti vectors in Fig.~\ref{fig:betti_numbers_toy_example}.
These Betti numbers are informative on the changing topology of the hypergraph snapshots in Fig.~\ref{fig:H_THG_snapshots}; however, they do not capture information on how the topology between the snapshots are related. For example, by observation of the hypergraph snapshots we know that there is one main component that persists through the entire sequence of ASCs, but this information can not be known directly from the Betti numbers. The Betti numbers do not tell the complete story of this component persisting the whole time. While they do tell us there is at least one component in each snapshot, these components do not necessarily need to be the same component in each snapshot to get the same Betti vectors. As such, we need to use a method to track how the homology is changing and related between the sequence of ASCs. To do this we implement zigzag persistent homology.

\subsection{Zigzag Persistent Homology} \label{ssec:ZZ_THG}
This section provides a conceptual introduction to persistent homology and how it generalizes to zigzag persistent homology. We suggest \cite{Otter2017,Munch2017} for a detailed introduction on persistent homology.

Persistent homology~\cite{Zomorodian2004}, a filtration tool from the field of Topological Data Analysis (TDA)~\cite{Edelsbrunner2002,Zomorodian2004}, is used to gain a sense of the shape and size of a dataset at multiple dimensions and filtration values. For example, it can measure connected components (dimension zero), holes (dimension one), voids (dimension two), and higher dimensional analogues, as well as an idea of their general size or geometry. 
Persistent homology measures these shapes using a parameterized filtration to detect when homology groups are born (appear) and die (disappear). 

To compute persistent homology a parameterization function is applied to the dataset to create a nested sequence of ASCs
\begin{equation}
K_{0} \subseteq K_{1} \subseteq K_{2} \subseteq \ldots \subseteq K_{n}.
\label{eq:nested_complexes}
\end{equation}
We can then calculate the homology of dimension $p$ for each complex, $H_p(K_{i})$, which is a vector space representing the $p$-dimensional structure of the space such as components, holes, voids, and higher dimensional features. 
However, this information does not yet yield how the homology of each ASC is related to the next ASC. 
To get this information, persistent homology uses the inclusions on the ASCs to induce linear maps on the vector spaces resulting in a construction called a persistence module $\mathcal{V}$:
\begin{equation}
H_p(K_{\alpha_0}) \hookrightarrow H_p(K_{\alpha_1}) \hookrightarrow H_p(K_{\alpha_2}) \hookrightarrow \ldots \hookrightarrow H_p(K_{\alpha_n}),
\label{eq:filtration}
\end{equation}
where $\hookrightarrow$ are the maps induced by the inclusion map between ASCs.
It should be noted that in the sequence of ASCs, each vertex must be unique and consistently identified.

The appearance and disappearance of classes at various dimensions in this object can be tracked, resulting in a summary known as a persistence barcode (alternatively a persistence diagram) $\mathcal{D} = \{D_0, D_1, \ldots, D_p\}$. 
For each homology generator which appears at $K_{b}$ and disappears at $K_{d}$, we draw an interval $[b,d]$ in the barcode. 
Taken together is the persistence barcode, which is the collection of persistence intervals (also called persistence pairs in the persistence diagram). 

\begin{figure}[h!] 
    \centering
    \begin{subfigure}[t]{.6\textwidth}
        \centering
        \includegraphics[width=1\textwidth]{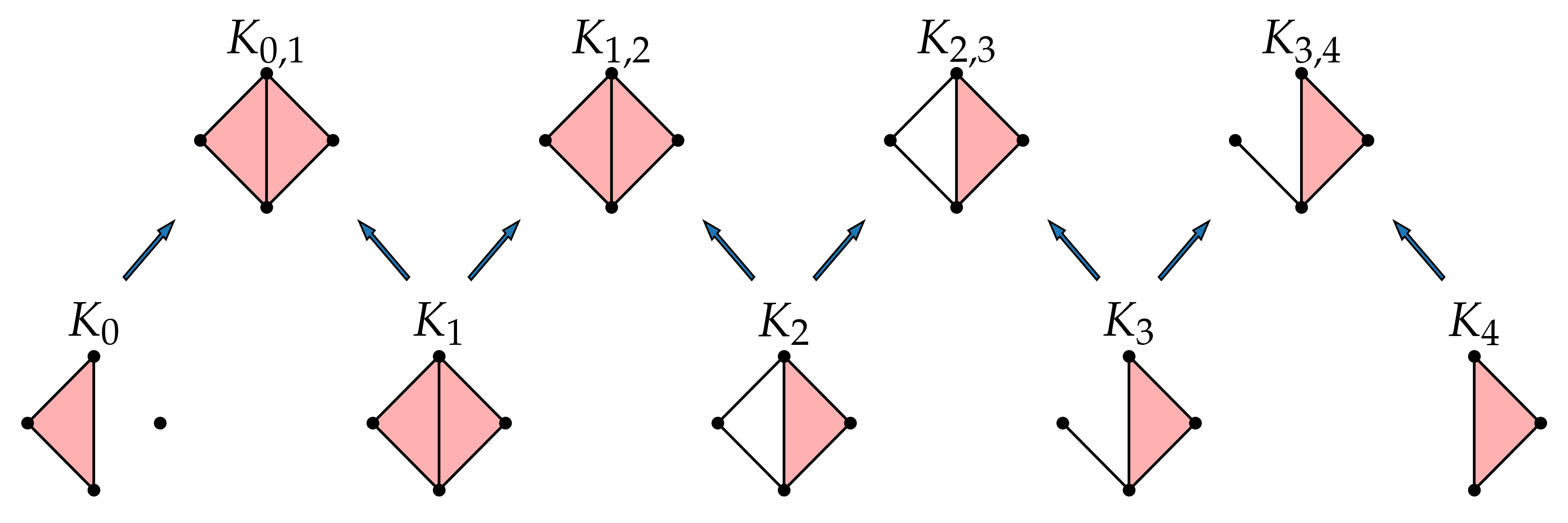}
        \caption{ASC sequence with unions.}
        \label{fig:K_THG_snapshots_minimal_example}
    \end{subfigure}
    \begin{subfigure}[t]{.3\textwidth}
        \centering
        \includegraphics[width=1\textwidth]{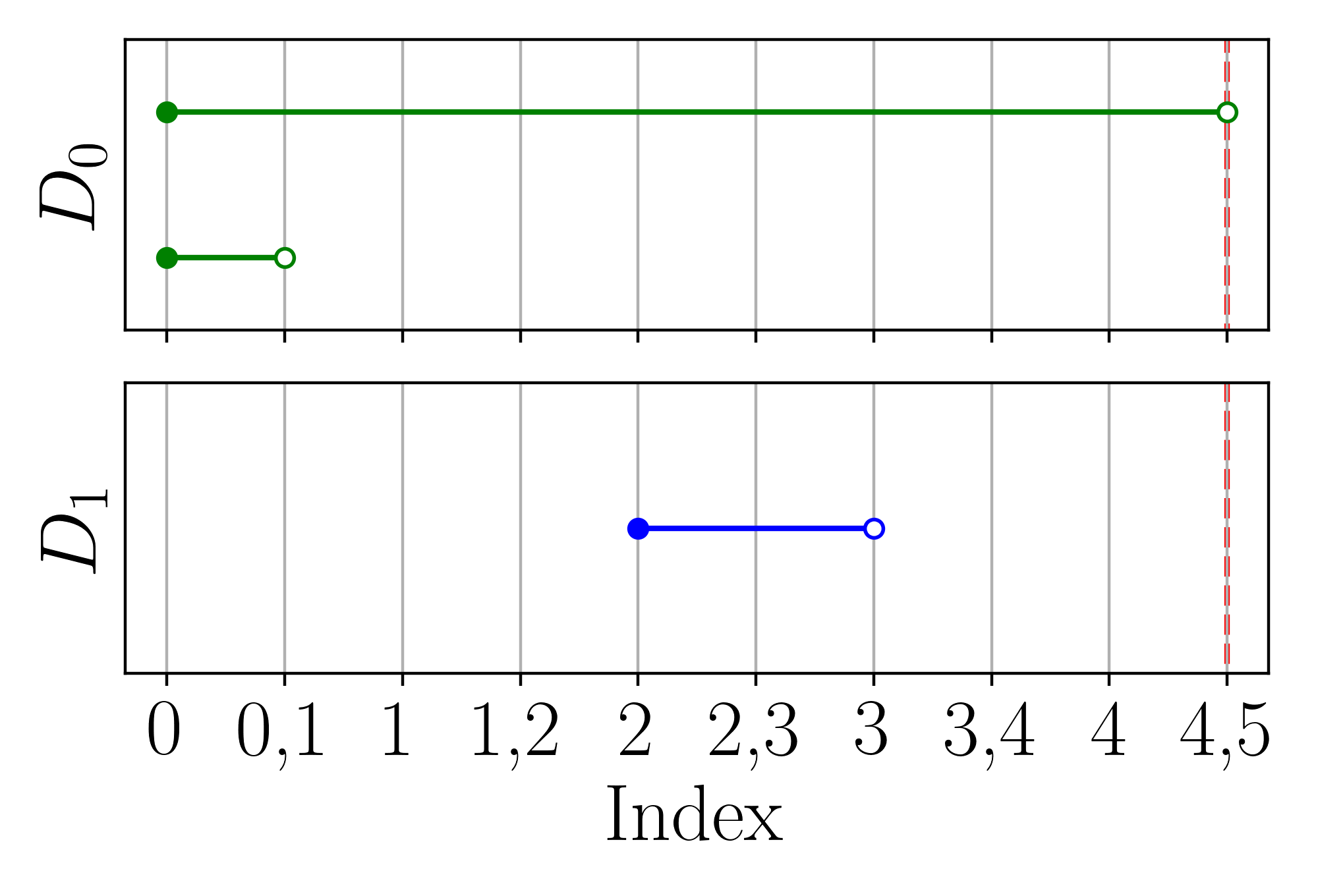}
        \caption{Zigzag persistence barcodes.}
        \label{fig:bar_code_example_minimal}
    \end{subfigure}
    \caption{Zigzag persistence module and resulting barcodes for dimensions 0 and 1 for toy example introduced in Fig.~\ref{fig:toy_example_THG}.}
    \label{fig:simple_zigzag_example}
\end{figure}

This persistent homology framework can be applied to study hypergraphs directly where a persistence module $\mathcal{V}$ is generated from a hypergraph, as described in~\cite{Ren2020}, by generating a sequence of subset ASC representations of a hypergraph.
However, a limitation of persistent homology is it requires each subsequent ASC to be a subset of the previous ASC to form the persistence module as shown in Eq.~\eqref{eq:nested_complexes}, which means at each step we are not allowed to remove simplices in the next ASC. 
There are many cases of real-world applications where we have a parameterized sequence of ASCs where simplices can both enter and exit the complex throughout the sequence.
To alleviate this issue zigzag persistence~\cite{Carlsson2010} can be applied, which allows for  arbitrary  subset directions in the ASC sequence:
\begin{equation}
    K_{0} \leftrightarrow K_{1} \leftrightarrow K_{2} \leftrightarrow \ldots \leftrightarrow K_{n},
\label{eq:bi_directional_zigzag_complexes}
\end{equation}
where $\leftrightarrow$ denotes one of the two inclusion maps: $\hookrightarrow$ or $\hookleftarrow$. 
A common special case of this definition is where the left and right inclusions alternate or zigzag. 
For most data analysis applications using zigzag persistent we artificially construction a sequence of ASCs taking this form by interweaving the original ASCs with either unions or intersections of adjacent ASCs. For example, in Fig.~\ref{fig:K_THG_snapshots_minimal_example} we use the union between the associated ASCs of the original hypergraph snapshots from Fig.~\ref{fig:H_THG_snapshots}. 
This sequence of interwoven ASCs fulfills the criteria of the zigzag inclusion map directions as 
\begin{equation}
K_0 
\hookrightarrow K_{0,1} 
\hookleftarrow K_{1} 
\hookrightarrow K_{1,2} 
\hookleftarrow K_{2} 
\hookrightarrow \ldots 
\hookleftarrow K_{{\ell-1}}  
\hookrightarrow K_{{\ell-1},{\ell}} 
\hookleftarrow K_{\ell}.
\label{eq:zigzag_intersection_ASC}
\end{equation}
for unions or 
\begin{equation}
K_0 
\hookleftarrow K_{0,1} 
\hookrightarrow K_{1} 
\hookleftarrow K_{1,2} 
\hookrightarrow K_{2} 
\hookleftarrow \ldots 
\hookrightarrow K_{{\ell-1}}  
\hookleftarrow K_{{\ell-1},{\ell}} 
\hookrightarrow K_{\ell}
\label{eq:zigzag_union_ASC}
\end{equation}
for intersections, where $K_{i,i+1} = K_i \cup K_{i+1}$. 

The inclusion maps are extended to linear maps between homology groups resulting in the zigzag persistence module tracking the changing homology of Eq.~\eqref{eq:zigzag_intersection_ASC} or \eqref{eq:zigzag_union_ASC} just as was the case for standard persistent homology.
Focusing on the case of the union, the zigzag persistent homology module is
\begin{equation}
H_p(K_{0}) \hookrightarrow H_p(K_{0,1})
\hookleftarrow H_p(K_{1})
\hookrightarrow H_p(K_{1,2})
\hookleftarrow H_p(K_{2})
\hookrightarrow \ldots 
\hookleftarrow H_p(K_{{n-1}})
\hookrightarrow H_p(K_{{n-1},{n}})
\hookleftarrow H_p(K_{n}).
\label{eq:zigzag_union_complexes}
\end{equation}

The same algebra leveraging the linear maps between homology groups to track persistence pairs for a standard filtration in persistent homology makes it possible to compute where (when) homology features are born and die based on the zigzag persistence module, however some of the intuition is lost.
Namely, we can again track the persistent homology using a persistence diagram $D = \{D_0, D_1, \ldots, D_p\}$ consisting of half-open intervals (persistence pairs) $[b, d)$; however, we now use the indices of the ASCs as the birth and death times instead of the filtration parameter. For example, if there is one-dimensional homology (i.e., a loop) that appears at  $K_2$ and persists until it disappears at $K_3$, we represent this as the persistence pair (2,3).
In the case of a class appearing or disappearing at the union (or intersection) complex $K_{{i},{i+1}}$, we use the half index pair ${i},{i+1}$.
If a topological feature persists in the last ASC in the zigzag persistence module we set its death past the last index with the pair $\ell, \ell+1$, where $\ell$ is the number of ASCs (without interwoven unions or intersections). 

To demonstrate how zigzag persistence tracks the changing topology in a sequence of ASCs we use a simple sequence of ASCs in Fig.~\ref{fig:K_THG__no_inter_snapshots}, which were derived from the toy example in Fig.~\ref{fig:toy_example_THG} using a sliding window procedure outlined in section~\ref{ssec:sliding_windows_for_HGs}. 
As a first example of the application of zigzag persistence to study temporal hypergraphs we return to our toy example. We used the unions between ASCs to get the ASCs shown as $[K_{0}, K_{0,1}, K_{1}, \ldots, K_{3,4}, K_{4}]$ in Fig.~\ref{fig:K_THG_snapshots_minimal_example} and the resulting zigzag persistence barcodes in Fig.~\ref{fig:bar_code_example_minimal}.
For this example we are only investigating the topological changes in dimensions 0 and 1 since there are no higher dimensional features.
There are two main changes in the homology of the ASCs that are captured in the persistence barcodes.
For dimension 0, we are tracking the connected components and how they relate. At $K_0$ we have two connected components (the 2-simplex as the triangle and 0-simplex as the point). As such, we set the birth of the two components at the index which they both appear: 0. Next, at $K_{0,1}$ the components combine as two conjoined 2-simplices. The joining of components forces one of the components to die while the other persists; the smaller of the two components dies (the 0-simplex) dies at the index $0,1$ with persistence interval $(0,(0,1))$ shown in the $D_0$ barcode of Fig.~\ref{fig:bar_code_example_minimal}. The combined component never separates or combines with another component again and therefor it persists for the remaining persistence module finally dying after $K_4$ or index $4,5$ (shown as the dashed red line) having the persistence interval $(0, (4,5))$ in $D_0$.
Moving to dimension 1, we are now interested in showing how the persistence barcode captures the formation and disappearance of loops in the persistence module. A loop is first formed in $K_2$ and persists until $K_3$. Therefor, this feature is represented as the persistence interval $(2, 3)$ in $D_1$ of Fig.~\ref{fig:bar_code_example_minimal}.
This example highlights how zigzag persistence captures changes in the topology of a sequence of ASCs.

\begin{figure}[h!] 
    \centering
    \begin{subfigure}[t]{0.69\textwidth}
        \centering
        \includegraphics[width=1\textwidth]{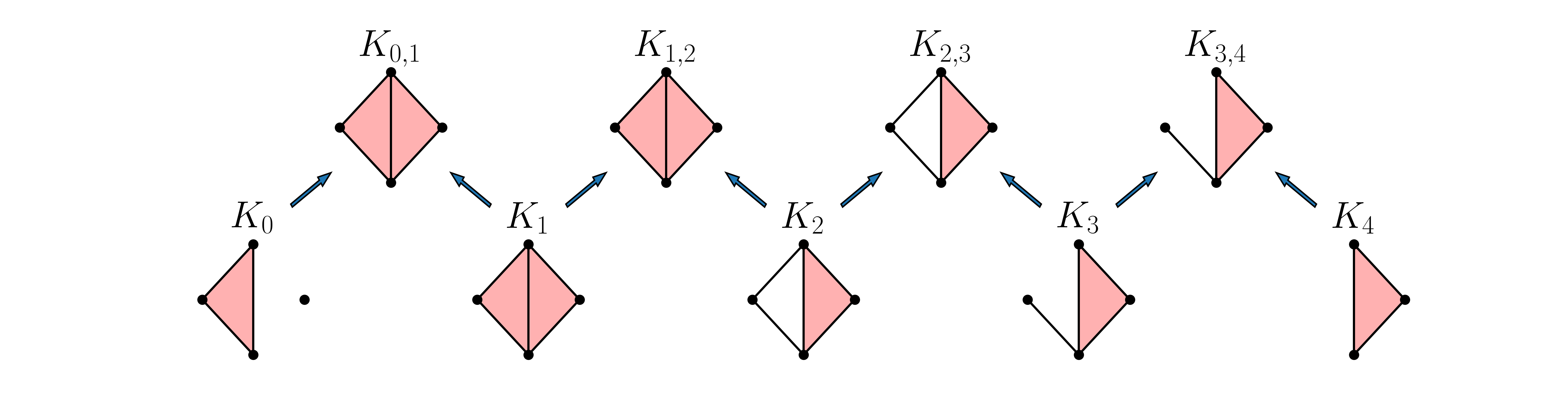}
        \caption{ASCs sequence with intermediate unions.}
        \label{fig:K_THG_snapshots_minimal_example_union}
    \end{subfigure}
    \begin{subfigure}[t]{0.3\textwidth}
        \centering
        \includegraphics[width=1\textwidth]{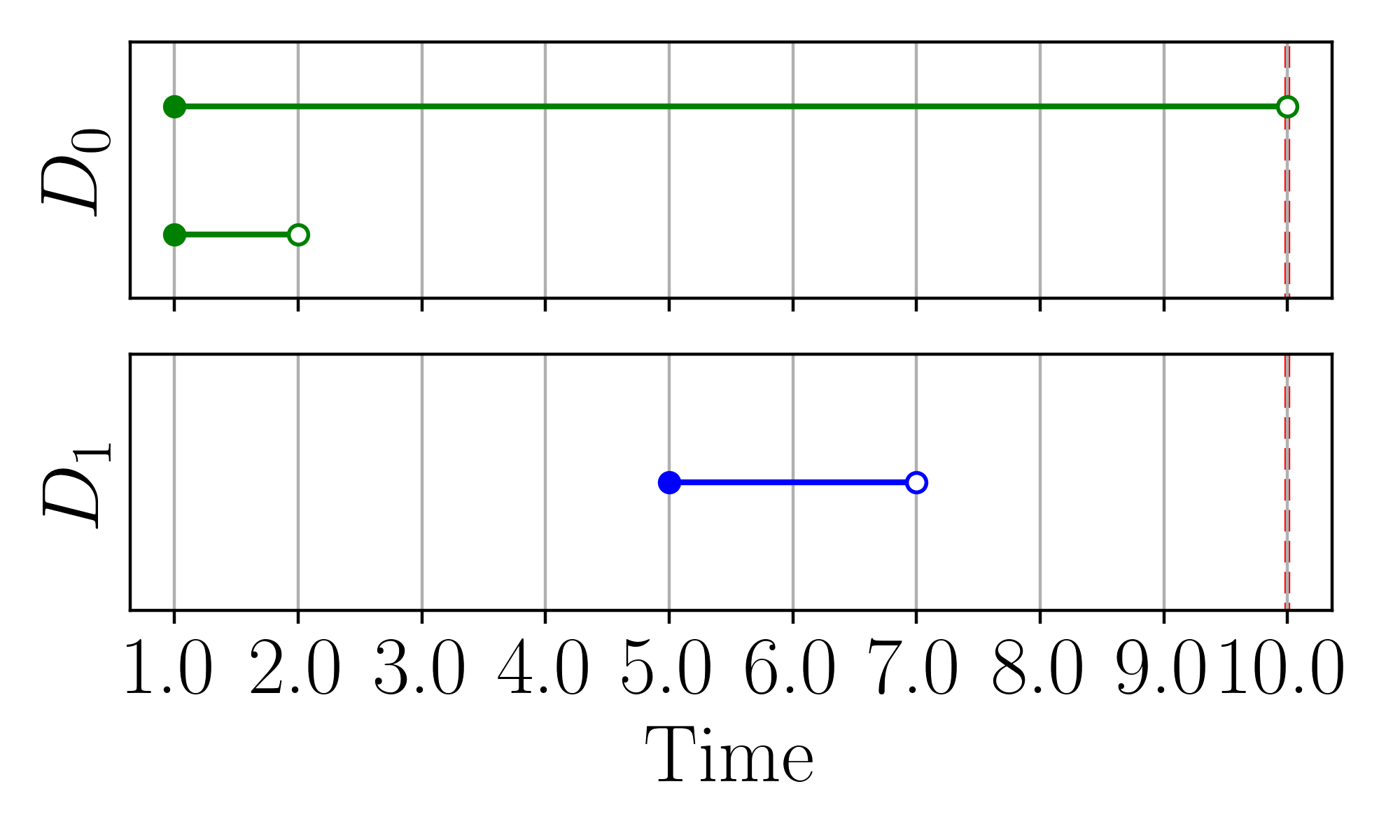}
        \caption{Barcodes using unions.}
        \label{fig:bar_code_example_minimal_windows_union}
    \end{subfigure}
    
    \begin{subfigure}[t]{0.69\textwidth}
        \centering
        \includegraphics[width=1\textwidth]{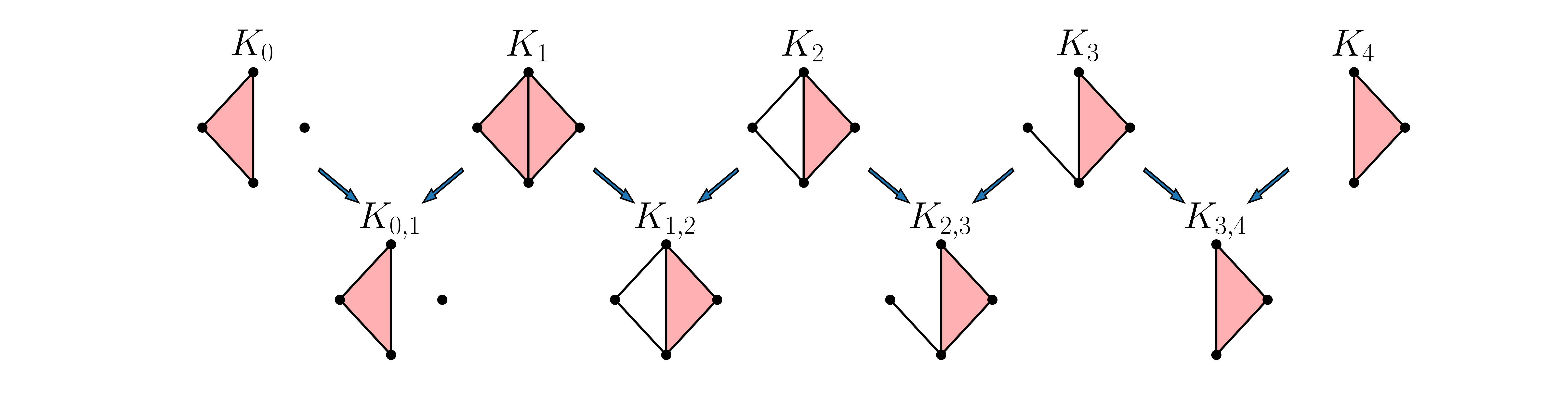}
        \caption{ASCs Sequence with intermediate intersections.}
        \label{fig:K_THG_snapshots_minimal_example_intersection}
    \end{subfigure}
    \begin{subfigure}[t]{0.3\textwidth}
        \centering
        \includegraphics[width=1\textwidth]{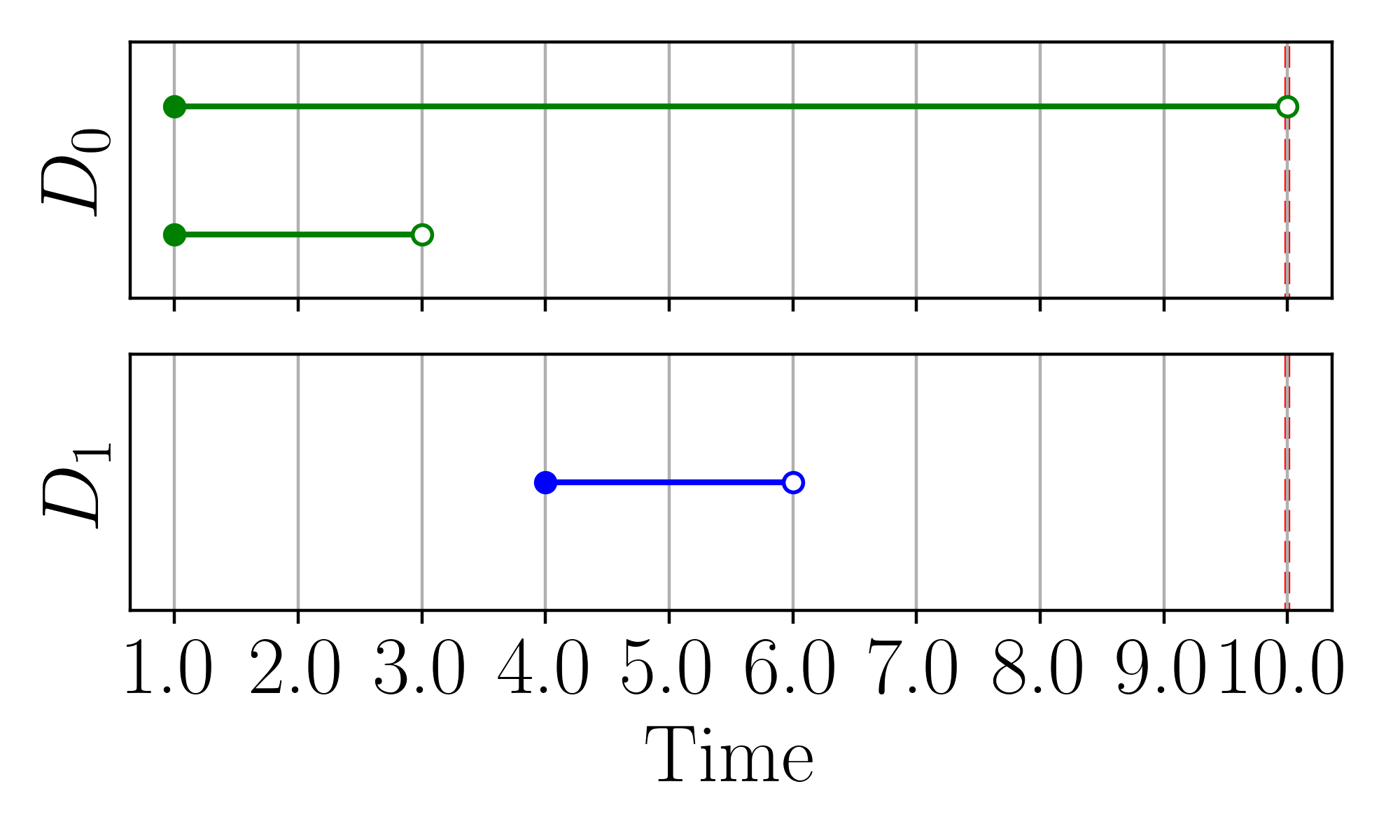}
        \caption{Barcodes using intersections.}
        \label{fig:bar_code_example_minimal_windows_intersection}
    \end{subfigure}
    \caption{Sequence of ASCs from the sliding window hypergraph snapshots for both union and intersections. Zigzag persistence barcodes for temporal hypergraph example with time associated ASCs.}
    \label{fig:ASC_windows_and_barcodes}
\end{figure}

In this work we are interested in the analysis of temporal hypergraphs, and as such we instead want to have the barcodes track the time at which homology appears and disappears instead of the indices. To do this we substitute the index for the average time of the window associated to each ASC as shown in Fig.~\ref{fig:ASC_windows_and_barcodes}. For the intermediate ASCs (unions or intersections) we use the average time of the two windows.
The only difference between the ASC sequence in Fig.~\ref{fig:bar_code_example_minimal} and Fig.~\ref{fig:bar_code_example_minimal_windows_union} is that Fig.~\ref{fig:bar_code_example_minimal_windows_union} has the times from the windows associated to the ASCs when computing the zigzag persistence. As such, the persistence barcode has time on the horizontal axis with the two intervals in $D_0$ and one in $D_1$ having the same sources (generators) as described in Fig.~\ref{fig:bar_code_example_minimal}. 

The resulting barcodes in Fig.~\ref{fig:ASC_windows_and_barcodes} shows that both the intersection and union methods for interweaving ASCs provide similar barcodes. We also found this same result when applying zigzag persistence to the data sets studied in this work. For the remainder of this work we will use the union method for studying temporal hypergraphs using zigzag persistence.

\section{Applications} \label{sec:results}

\subsection{Social Network Analysis}
\label{subsec:SNA}
To demonstrate the functionality of analyzing temporal hypergraph data through zigzag persistence we use Reddit data with COVID-related subreddits. This data is known as the PAPERCRANE dataset~\cite{Papercrane2022}.

The dataset subset we use spans from 1/20/20 to 3/31/20. This section captures the initial formation of the subreddits during the onset of COVID-19.
The active subreddits related to COVID-19 in the dataset during this time are listed in Table~\ref{tab:subreddit_data} with summary statistics on the number of threads and authors.

\begin{table}[h!]
    \caption{Subreddits related to covid from the PAPERCRANE datatset with number of threads and authors of each subreddit}
    \centering
    \begin{tabular}{lccc}
        Subreddit            & Active Dates        & Threads  & Authors \\ \hline
        CCP\_virus           & 3/27 - 3/31         & 169      & 79     \\
        COVID19              & 2/15 - 3/31         & 8668     & 22020    \\
        COVID19positive      & 3/13 - 3/31         & 1462     & 6682    \\
        China\_Flu           & 1/20 - 3/31         & 55466    & 62944    \\
        Coronavirus          & 1/20 - 3/31         & 153025   & 396427   \\
        CoronavirusCA        & 3/01  - 3/31        & 2930     & 5370    \\
        CoronavirusRecession & 3/19 - 3/31         & 1574     & 6548    \\
        CoronavirusUK        & 2/15 - 3/31         & 8654     & 10230    \\
        CoronavirusUS        & 2/16 - 3/31         & 18867    & 29809    \\
        Covid2019            & 2/16 - 3/31         & 2437     & 1531     \\
        cvnews               & 1/25 - 3/31         & 4233     & 2181     \\
        nCoV                 & 1/20 - 3/31         & 3949     & 1902     \\ \hline
    \end{tabular}
    \label{tab:subreddit_data}
\end{table}

\begin{figure}[h!] 
    \centering
    \includegraphics[width=1\textwidth]{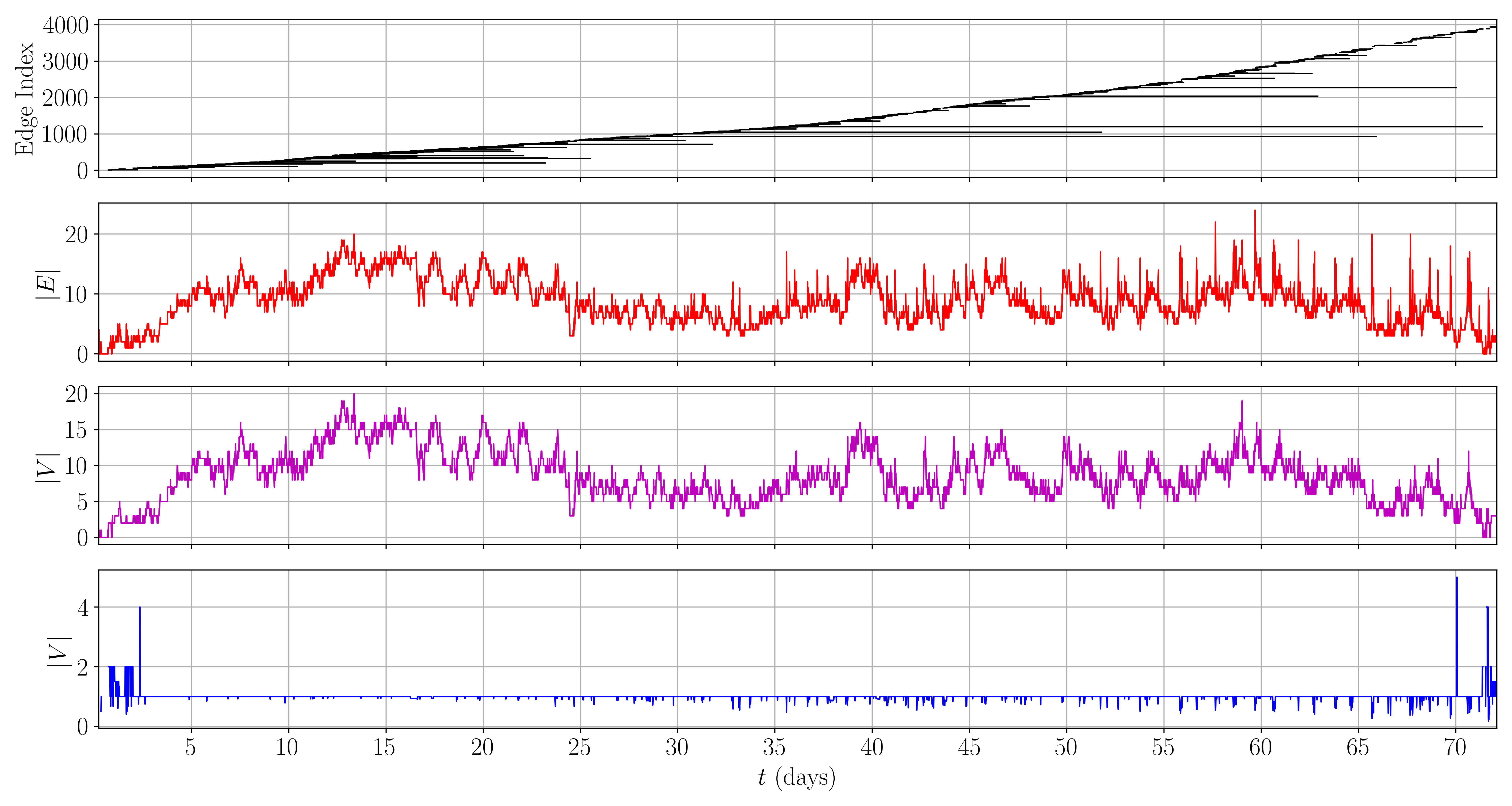}
    \caption{Summary statistics for size of temporal hypergraph snapshots. The top is the interval associated to each edge (sorted by start time), the middle figure is the number of edges in the hypergraph snapshots, and the bottom figure is the number of vertices in the hypergraph snapshots.}
    \label{fig:summary_statistics_reddit_nCoV}
\end{figure}

In this analysis we only use the nCoV subreddit due to its manageable size and interpretability.
The temporal intervals for the edges are constructed from the author interaction information. We construct edge intervals based on the first and last times an author posted in each thread. These intervals are visualized in the top subfigure of Fig.~\ref{fig:summary_statistics_reddit_nCoV}.

We set the window size of 1 hour with a shift of 15 minutes. 
This window size captures the necessary granularity to see changes in the dynamics of the subreddit. Applying this sliding window results in 6899 windows. 
The number of nodes and edges of each hypergraph snapshot is shown in Fig.~\ref{fig:summary_statistics_reddit_nCoV}. This initial data exploration shows that the size of the subreddit initially increases to a peak popularity at approximately two weeks into the subreddit or day 14. After this, the size steadily decreases. The edge intervals in the top subfigure of Fig.~\ref{fig:summary_statistics_reddit_nCoV} shows that the majority of intervals are very short, while a few exhibit long intervals lasting as long as 38 days. This initial exploration does not capture how the shape of the underlying hypergraph snapshots is evolving.

There are many questions about the underlying network that can not be directly answered from these simple summary statistics. For example, is each thread dominated by unique authors or do many threads share users? Is the social network dense, centralized, fragmented? Do any of these characteristics change over time?

\begin{figure}[h!] 
    \centering
    \includegraphics[width=0.85\textwidth]{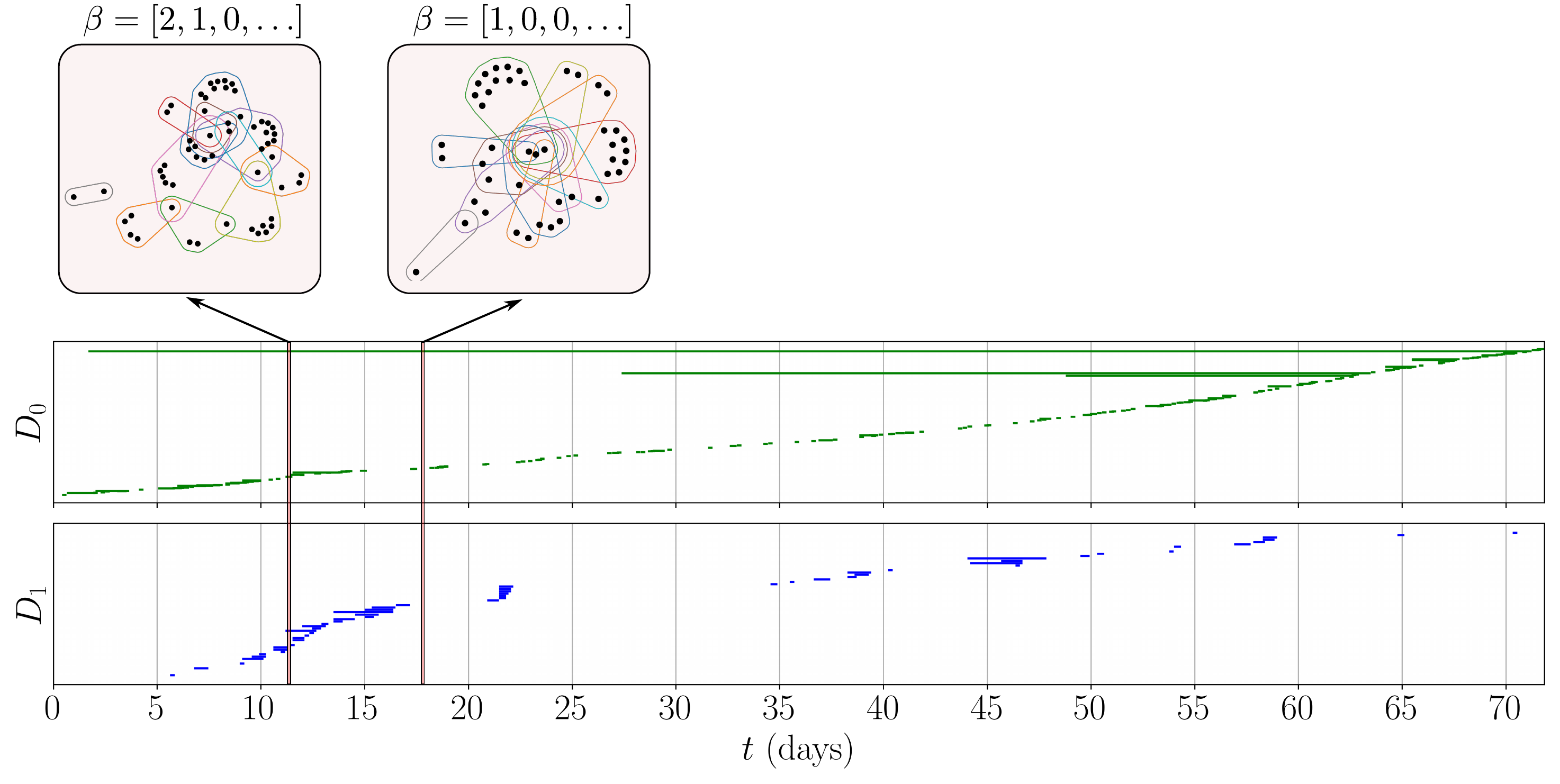}
    \caption{Zigzag Persistence of temporal hypergraph representation of the CCP\_virus subreddit with example hypergraph snapshot and associated ASC.}
    \label{fig:nCoV_ZZ_persistence_union}
\end{figure}

Understanding the topological summary of the hypergraph snapshots is important to understand the type of communication that is occurring. For example, many one-dimensional homology features are representative of disconnected conversations of holes in the communication structure. However, this could be captured just using the Betti sequence at each snapshot. What the zigzag persistence also captures is the severity of the holes based on their longevity. Consider a hole in communication that persists for several days. This could be representative of a lack of information communication throughout the community.
These summary statistics additionally do not provide any information on how the threads in the subreddit are related and their longevity. Using zigzag persistence we can capture information about the longevity of communications using the zero-dimensional homology. A long interval in the zero-dimensional zigzag persistence barcode is representative of a conversation persisting over a long period of time. 
In Fig.~\ref{fig:nCoV_ZZ_persistence_union} are the resulting zigzag persistence barcodes using the union between the associated ASCs of the hypergraph snapshots. 

First, we see that we can capture how fragmented the social network is with one main component shown in the zero-dimensional barcode that persists for almost the entire duration of the subreddit. Additionally, the short intervals in dimension zero are characteristic of other side conversations, which either split from or merged into the main conversation or were entirely separate conversations. An example of one of these conversations is shown in the hypergraph snapshot at day 10 in Fig.~\ref{fig:nCoV_ZZ_persistence_union} where the main component is composed of all of the threads with exception to one thread between just two authors. Having the main component suggests that many of the threads in the subreddit share at least one common author between threads.

We can also demonstrate that the network shows a change in its centralization over time. Specifically, during regions where many $D_1$ persistence intervals are present we know that the network has several loops, which are characteristic of non-centralized social networks. These changes from centralized to non-centralized social hypergraph snapshots are likely due to the number of authors active and a bifurcation of social network dynamics. For example, in the snapshot at day 10 in Fig.~\ref{fig:nCoV_ZZ_persistence_union} there is a main loop in the main component of the hypergraph snapshot captured, and the main component does not have a clearly centralized structure. However, approximately one week later at day 18, there is a clearly centralized structure to the hypergraph which has no one-dimensional features. With both a low number of (or no) one-dimensional features and only one component, the zigzag persistence can give insight into the centralization of the hypergraph and underlying social network. 

\subsection{Cyber Data Analysis}
\label{subsec:CDA}
For the analysis of cyber data we use the Operationally Transparent Cyber dataset (OpTC)~\cite{OpTC} created by the Defense Advanced Research Projects Agency (DARPA).
This dataset consists of network and host logging from hundreds of windows hosts over a week period. The dataset consists of two groups of user activity: benign and malicious. The malicious activity occurs over a three day period in which several attacks are executed.

Our goal is to analyze demonstrate how these attacks show up in the zigzag persistence barcodes for a hypergraph formation from the data log.
The data log is composed of 64 columns describing each action in the network.
In this section we only use the timestamps, src\_ip, image\_path, and dest\_port, as these are needed to construct the temporal hypergraph representation of the data we study using zigzag persistence.

We construct hypergraph snapshots by again using a sliding window procedure, but now the intervals associated to each edge are only time points as the cyber data only has the time stamp at which the action occurred. We used a sliding window with width $w = 30$ minutes and shift $s = 5$ minutes. We chose this window size based on the duration of malicious activity lasting for approximately 2 hours with 30 minute windows being fine grained enough to capture the transition from benign to malicious.

To demonstrate how zigzag persistence can detect a cyber attack we will look at two instances of malicious activity on two different hosts.
Namely, we investigate two cases of a cyber attack; the first on 9/23/19 from red agent LUAVR71T with source IP 142.20.56.202 on host 201 and the second on 9/24/19 from agent 4BW2MKUF with source IP 142.20.57.246 on host 501.
The first sequence of attack beings at approximately 11:23 to 13:24 on 9/23/19 and the second sequence from approximately 10:46 to 13:11.

The hypergraphs were constructed using the destination ports as the hyperedges and the image paths as nodes. This formation captures the structure of the cyber data in the sense that the destination ports as hyperedges capture the relation between the actions (image paths) used. Additionally, we only use a subset of the full data for a single source IP. By only looking at this sub-hypergraph we capture information about the specific agent associated to the source IP.

\begin{figure}[h!] 
    \centering
    \begin{subfigure}[t]{0.75\textwidth}
        \centering
        \includegraphics[width=1\textwidth]{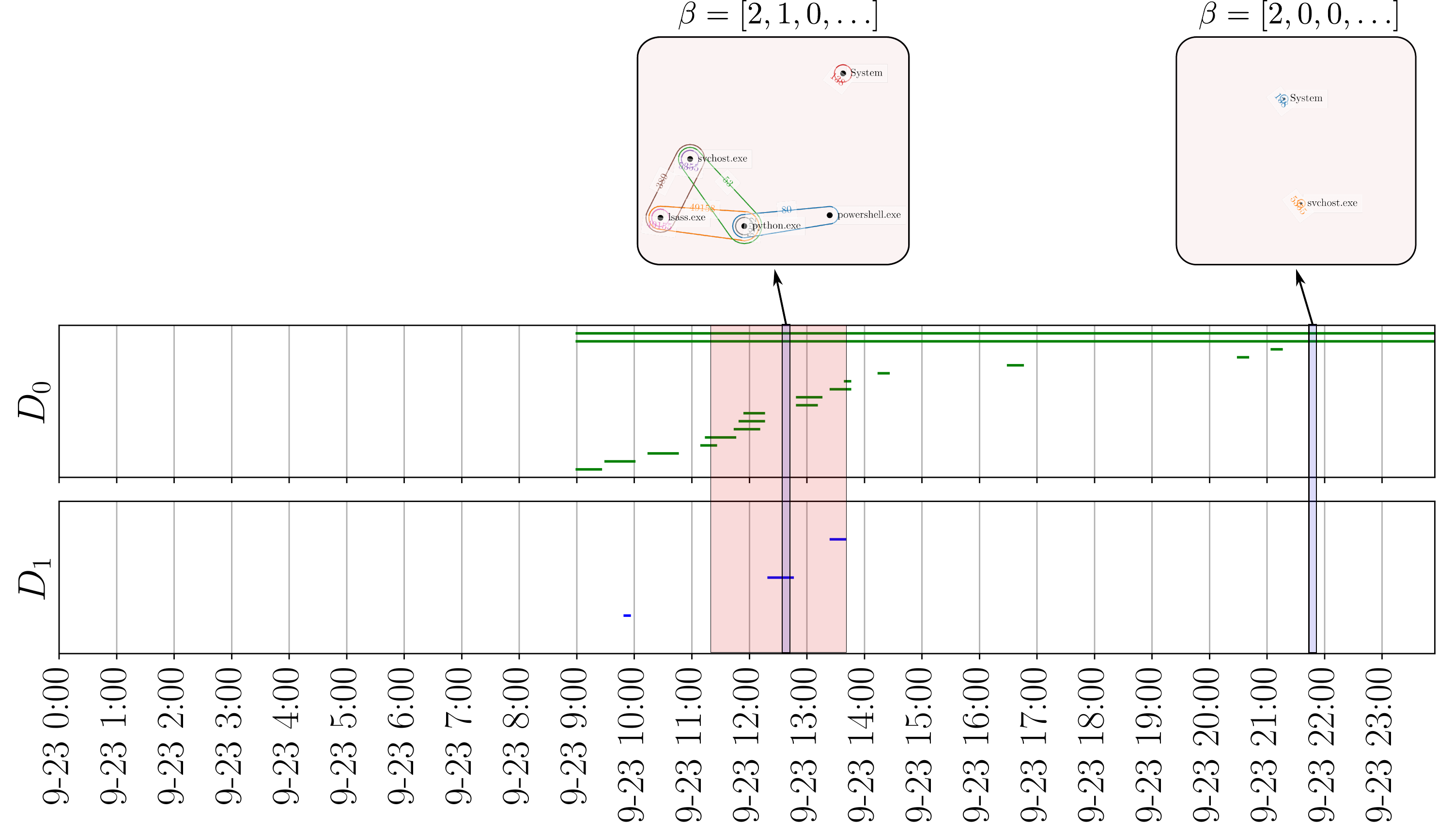}
        \caption{Zigzag persistence barcodes of red team agent LUAVR71T with source IP 142.20.56.202 on host 201 for the day of 9/23/19.}
        \label{fig:OpTC_201_23_ZZ_barcodes}
    \end{subfigure}
    \par\bigskip
    \begin{subfigure}[t]{0.75\textwidth}
        \centering
        \includegraphics[width=1\textwidth]{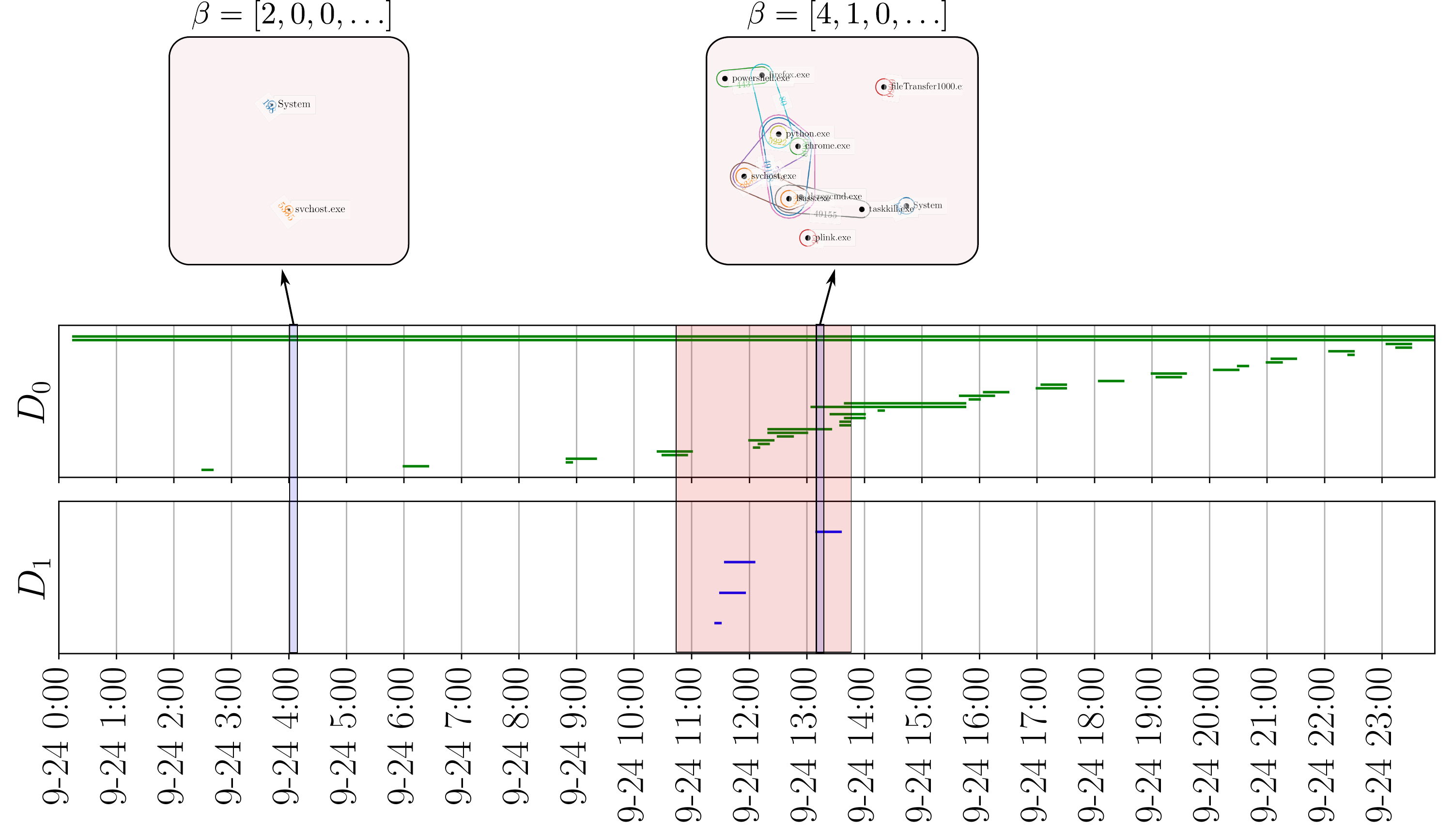}
        \caption{Zigzag persistence barcodes of red team agent 4BW2MKUF with source IP 142.20.57.246 on host 501 for the day of 9/24/19.}
        \label{fig:OpTC_501_24_ZZ_barcodes}
    \end{subfigure}
    \caption{Zigzag persistence barcodes with example hypergraphs at two windows for OpTC data during an attack on the 23rd and 24th. The region highlighted in red is the time the red team agent was activity attacking.}
    \label{fig:OpTC_zigzag_persistence_barcodes}
\end{figure}

The zigzag persistence barcodes associated the the destination port/image path hypergraph snapshots for the first sequence of attacks are shown in Fig.~\ref{fig:OpTC_201_23_ZZ_barcodes}. Before 9:00 there was no cyber activity and as such no barcodes during that period. The region highlighted in red from 11:23 to 13:24 is the active time of the cyber attacks.
During this region we highlight a specific hypergraph for the window starting at approximately 12:35 which is exemplary of malicious activity. Additionally, at approximately 21:50, we show another exemplary window on standard benign activity. During this activity there are typically only two singletons which persist over time.
A similar pair of hypergraphs for malicious and benign activity are shown in the second sequence of malicious activity on 9/24/19.
However, what is not captured by the snapshots are the dynamics and quickly changing topology of the snapshots during malicious activity and relatively stationary dynamics and simple topology during benign activity.

Zigzag persistence is able to capture the changing dynamics and topology that is characteristic of malicious cyber activity. This is shown in both the barcodes for $D_0$ and $D_1$ for both sequences of malicious activity as shown in Fig.~\ref{fig:OpTC_zigzag_persistence_barcodes}. Specifically, during malicious activity there tends to be more, short-lived persistence pairs in $D_0$ and the appearance of one-dimensional homology in $D_1$. In comparison, during benign activity, there is little to no one-dimensional homology and little change in the number of components captured through $D_0$.
%!TEX root = ..\main.tex
%-------------------------------
%*******************************

\section{Conclusion} \label{sec:conclusion}

In this work we developed an implementation of zigzag persistence for studying temporal hypergraphs.
To demonstrate the functionality of our method we apply it to study both social network and cyber security data represented as temporal hypergraphs.
For the social network analysis we were able to show how the resulting zigzag persistence barcodes capture the dynamics of the temporal hypergraphs topology which captures information about the changing centrality of the hypergraphs through $D_1$. Furthermore, we show that the conversation is composed of one main component that persists over the entire time period of the social network we studied.
When studying the cyber data we found that we were able to detect malicious from benign activity with zigzag persistence. During malicious activity we showed that there tends to be persistence pairs in $D_1$ as well as more persistence pairs in $D_0$ in comparison to during benign activity.

Future work for this method includes an investigation of vectorization techniques of the zigzag persistence diagrams for automating cyber security analysis. We also plan to study how we can leverage the temporal hypergraph representation and zigzag persistence for detecting bot activity in social network data.

\bibliographystyle{splncs04}
\bibliography{main}

\begin{thebibliography}{10}
\providecommand{\url}[1]{\texttt{#1}}
\providecommand{\urlprefix}{URL }
\providecommand{\doi}[1]{https://doi.org/#1}

\bibitem{Adams2017}
Adams, H., Emerson, T., Kirby, M., Neville, R., Peterson, C., Shipman, P.,
  Chepushtanova, S., Hanson, E., Motta, F., Ziegelmeier, L.: Persistence
  images: A stable vector representation of persistent homology. Journal of
  Machine Learning Research  \textbf{18}(8),  1--35 (2017),
  \url{http://jmlr.org/papers/v18/16-337.html}

\bibitem{OpTC}
Agency, D.A.R.P.: Operationally transparent cyber (optc) data release (2020)

\bibitem{Aktas2019}
Aktas, M.E., Akbas, E., Fatmaoui, A.E.: Persistence homology of networks:
  methods and applications. Applied Network Science  \textbf{4}(1) (Aug 2019).
  \doi{10.1007/s41109-019-0179-3}

\bibitem{Amzquita2020}
Am{\'{e}}zquita, E.J., Quigley, M.Y., Ophelders, T., Munch, E., Chitwood, D.H.:
  The shape of things to come: Topological data analysis and biology, from
  molecules to organisms. Developmental Dynamics  \textbf{249}(7),  816--833
  (Apr 2020). \doi{10.1002/dvdy.175}

\bibitem{Papercrane2022}
Baumgartner, J., Zannettou, S., Keegan, B., Squire, M., Blackburn, J.: The
  pushshift reddit dataset. PUSHSHIFT (jan 2020).
  \doi{https://doi.org/10.5281/zenodo.3608135}, reddit-hazelnut prepared for
  the Social Network ProblemShop (Jan 24-Feb 4, 2022). Ottawa, Canada.
  Derivative of Reddit data obtained via pushshift.io API for the period
  January 1, 2019 to February 28, 2019.

\bibitem{Bubenik2015}
Bubenik, P.: Statistical topological data analysis using persistence
  landscapes. Journal of Machine Learning Research  \textbf{16}(3),  77--102
  (2015), \url{http://jmlr.org/papers/v16/bubenik15a.html}

\bibitem{Carlsson2010}
Carlsson, G., de~Silva, V.: Zigzag persistence. Foundations of Computational
  Mathematics  \textbf{10}(4),  367--405 (Apr 2010).
  \doi{10.1007/s10208-010-9066-0}

\bibitem{Cencetti2021}
Cencetti, G., Battiston, F., Lepri, B., Karsai, M.: Temporal properties of
  higher-order interactions in social networks. Scientific Reports
  \textbf{11}(1) (Mar 2021). \doi{10.1038/s41598-021-86469-8}

\bibitem{DavidBoyce2012}
David~Boyce, B.R.: Modeling Dynamic Transportation Networks. Springer Berlin
  Heidelberg (2012)

\bibitem{Edelsbrunner2002}
Edelsbrunner, Letscher, Zomorodian: Topological persistence and simplification.
  Discrete {\&}amp; Computational Geometry  \textbf{28}(4),  511--533 (Nov
  2002). \doi{10.1007/s00454-002-2885-2}

\bibitem{Estrada2006}
Estrada, E., Rodr{\'{\i}}guez-Vel{\'{a}}zquez, J.A.: Subgraph centrality and
  clustering in complex hyper-networks. Physica A: Statistical Mechanics and
  its Applications  \textbf{364},  581--594 (May 2006).
  \doi{10.1016/j.physa.2005.12.002}

\bibitem{Feng2021}
Feng, S., Heath, E., Jefferson, B., Joslyn, C., Kvinge, H., Mitchell, H.D.,
  Praggastis, B., Eisfeld, A.J., Sims, A.C., Thackray, L.B., Fan, S., Walters,
  K.B., Halfmann, P.J., Westhoff-Smith, D., Tan, Q., Menachery, V.D., Sheahan,
  T.P., Cockrell, A.S., Kocher, J.F., Stratton, K.G., Heller, N.C., Bramer,
  L.M., Diamond, M.S., Baric, R.S., Waters, K.M., Kawaoka, Y., McDermott, J.E.,
  Purvine, E.: Hypergraph models of biological networks to identify genes
  critical to pathogenic viral response. {BMC} Bioinformatics  \textbf{22}(1)
  (May 2021). \doi{10.1186/s12859-021-04197-2}

\bibitem{Fischer2021}
Fischer, M.T., Arya, D., Streeb, D., Seebacher, D., Keim, D.A., Worring, M.:
  Visual analytics for temporal hypergraph model exploration. {IEEE}
  Transactions on Visualization and Computer Graphics  \textbf{27}(2),
  550--560 (Feb 2021). \doi{10.1109/tvcg.2020.3030408}

\bibitem{Gasparovic2021}
Gasparovic, E., Gommel, M., Purvine, E., Sazdanovic, R., Wang, B., Wang, Y.,
  Ziegelmeier, L.: Homology of graphs and hypergraphs (May 2021),
  \url{https://www.youtube.com/watch?v=XeNBysFcwOw}

\bibitem{golczynski2021end}
Golczynski, A., Emanuello, J.A.: End-to-end anomaly detection for identifying
  malicious cyber behavior through nlp-based log embeddings. arXiv preprint
  arXiv:2108.12276  (2021)

\bibitem{hanselmann2020canet}
Hanselmann, M., Strauss, T., Dormann, K., Ulmer, H.: Canet: An unsupervised
  intrusion detection system for high dimensional can bus data. Ieee Access
  \textbf{8},  58194--58205 (2020)

\bibitem{Harary1997}
Harary, F., Gupta, G.: Dynamic graph models. Mathematical and Computer
  Modelling  \textbf{25}(7),  79--87 (Apr 1997).
  \doi{10.1016/s0895-7177(97)00050-2}

\bibitem{Husein2019}
Husein, I., Mawengkang, H., Suwilo, S., Mardiningsih: Modeling the transmission
  of infectious disease in a dynamic network. Journal of Physics: Conference
  Series  \textbf{1255}(1),  012052 (aug 2019).
  \doi{10.1088/1742-6596/1255/1/012052}

\bibitem{Joslyn2021}
Joslyn, C.A., Aksoy, S.G., Callahan, T.J., Hunter, L.E., Jefferson, B.,
  Praggastis, B., Purvine, E., Tripodi, I.J.: Hypernetwork science: From
  multidimensional networks to~computational topology. In: Unifying Themes in
  Complex Systems X, pp. 377--392. Springer International Publishing (2021).
  \doi{10.1007/978-3-030-67318-5_25}

\bibitem{Khasawneh2016}
Khasawneh, F., Munch, E.: Chatter detection in turning using persistent
  homology. Mechanical Systems and Signal Processing  \textbf{70-71},  527 --
  541 (2016). \doi{10.1016/j.ymssp.2015.09.046}

\bibitem{Munch2017}
Munch, E.: A user's guide to topological data analysis. Journal of Learning
  Analytics  \textbf{4}(2) (Jul 2017). \doi{10.18608/jla.2017.42.6}

\bibitem{Myers2019}
Myers, A., Munch, E., Khasawneh, F.A.: Persistent homology of complex networks
  for dynamic state detection. Physical Review E  \textbf{100}(2) (Aug 2019).
  \doi{10.1103/physreve.100.022314}

\bibitem{Myers2022a}
Myers, A., Muñoz, D., Khasawneh, F., Munch, E.: Temporal network analysis
  using zigzag persistence (2022)

\bibitem{Otter2017}
Otter, N., Porter, M.A., Tillmann, U., Grindrod, P., Harrington, H.A.: A
  roadmap for the computation of persistent homology. {EPJ} Data Science
  \textbf{6}(1) (Aug 2017). \doi{10.1140/epjds/s13688-017-0109-5}

\bibitem{Ren2020}
Ren, S.: Persistent homology for hypergraphs and computational tools
  {\textemdash} a survey for users. Journal of Knot Theory and Its
  Ramifications  \textbf{29}(13),  2043007 (Nov 2020).
  \doi{10.1142/s0218216520430075}

\bibitem{Schaefer2018}
Schäfer, B., Witthaut, D., Timme, M., Latora, V.: Dynamically induced
  cascading failures in power grids. Nature Communications  \textbf{9}(1) (may
  2018). \doi{10.1038/s41467-018-04287-5}

\bibitem{Skaf2022}
Skaf, Y., Laubenbacher, R.: Topological data analysis in biomedicine: A review.
  Journal of Biomedical Informatics  \textbf{130},  104082 (Jun 2022).
  \doi{10.1016/j.jbi.2022.104082}

\bibitem{Skyrms2000}
Skyrms, B., Pemantle, R.: A dynamic model of social network formation.
  Proceedings of the National Academy of Sciences  \textbf{97}(16),  9340--9346
  (aug 2000). \doi{10.1073/pnas.97.16.9340}

\bibitem{Tempelman2020}
Tempelman, J.R., Khasawneh, F.A.: A look into chaos detection through
  topological data analysis. Physica D: Nonlinear Phenomena  \textbf{406},
  132446 (May 2020). \doi{10.1016/j.physd.2020.132446}

\bibitem{Tymochko2020}
Tymochko, S., Munch, E., Khasawneh, F.: Using zigzag persistent homology to
  detect hopf bifurcations in dynamical systems. Algorithms  \textbf{13}(11)
  (Oct 2020). \doi{10.3390/a13110278}

\bibitem{Xu2019b}
Xu, M., Radhakrishnan, S., Kamarthi, S., Jin, X.: Resiliency of mutualistic
  supplier-manufacturer networks. Scientific Reports  \textbf{9}(1) (sep 2019).
  \doi{10.1038/s41598-019-49932-1}

\bibitem{Yesilli2022}
Yesilli, M.C., Chumley, M.M., Chen, J., Khasawneh, F.A., Guo, Y.: Exploring
  surface texture quantification in piezo vibration striking treatment ({PVST})
  using topological measures. In: Volume 2: Manufacturing Processes;
  Manufacturing Systems. American Society of Mechanical Engineers (Jun 2022).
  \doi{10.1115/msec2022-86659}

\bibitem{Zomorodian2004}
Zomorodian, A., Carlsson, G.: Computing persistent homology. Discrete {\&}amp;
  Computational Geometry  \textbf{33}(2),  249--274 (Nov 2004).
  \doi{10.1007/s00454-004-1146-y}

\end{thebibliography}

%\printbibliography

\end{document}